# Jet conversion photons from an anisotropic *Quark-Gluon-Plasma*

Lusaka Bhattacharya[a] and Pradip Roy[b]

*Saha Institute of Nuclear Physics*
*1/AF Bidhannagar, Kolkata - 700064, INDIA*

## ABSTRACT

We calculate the $p_T$ distributions of jet conversion photons from *Quark Gluon Plasma* with pre-equilibrium momentum-space anisotropy. A phenomenological model has been used for the time evolution of hard momentum scale $p_{\text{hard}}(\tau)$ and anisotropy parameter $\xi(\tau)$. As a result of pre-equilibrium momentum-space anisotropy, we find significant modification of the jet conversion photon $p_T$ distribution. For example, with *fixed initial condition* (FIC) pre-equilibrium anisotropy, we predict significant enhancement of the jet-photon $p_T$ distribution in the entire region, whereas for pre-equilibrium anisotropy with *fixed final multiplicity* (FFM), suppression of the jet conversion photons $p_T$ distribution is observed. The results with FFM (as it is the most realistic situation) have been compared with high $p_T$ PHENIX photon data. It is found that the data is reproduced well if the isotropization time lies within 1.5 fm/c.

## 1 Introduction

The main goal of experiments which perform ultra-relativistic heavy-ion collisions is to produce and study the properties of a deconfined plasma of quarks and gluons. This new state of matter, called quark-gluon plasma (QGP) is expected to be formed at temperature of the oder of $170 - 200$ MeV. Relativistic heavy ion colliders (RHIC), at Brookhaven National Laboratory and Large Hadron Collider (LHC) at CERN, are designed to produce and study strongly interacting matter at high temperature and/or density. The possibility of QGP formation at RHIC energies, is supported by the observation of high $p_T$ hadron suppression in the central $A-A$ collisions compared to binary scaled $p-p$ collisions [1]. This observation is referred to as *jet-quenching*. Another most important task is to characterize different properties of this new state of matter, such as isotropization/thermalization.

The most difficult problem lies in the determination of isotropization and thermalization time scales ($\tau_{\text{iso}}$ and $\tau_{\text{therm}}$) [3]. Studies on elliptic flow (upto about $p_T \sim 1.5 - 2$ GeV) using ideal hydrodynamics indicate that the matter produced in such collisions becomes isotropic with $\tau_{\text{iso}} \sim 0.6$ fm/c [2, 3, 4]. On the contrary, perturbative estimates yield much slower thermalization of QGP [35]. However, recent hydrodynamical studies [6] have shown that due to the poor knowledge of the initial conditions, there is a sizable amount of uncertainty in the estimation of thermalization or isotropization time. The other uncertain parameters are the transition temperature $T_c$, the spatial profile, and the effects of flow. Thus it is necessary to find suitable probes which are sensitive to these parameters. Electromagnetic probes have long been considered to be one of the most promising tools to characterize the initial state of the collisions [7, 8]. Because of the very nature of their interactions with the constituents of the system they tend to leave the system without much change of their energy and momentum. Photons (dilepton as well) can be one such observables.

---

[a]E-mail address: lusaka.bhattacharya@saha.ac.in
[b]E-mail address: pradipk.roy@saha.ac.in
[3]From now on we will concentrate on the most simplest possibility that is both the time-scale are the same, $\tau_{\text{therm}} = \tau_{\text{iso}}$.

But, photons can carry information about the plasma initial conditions [9, 10, 11] only if the observed flow effects from the late stages of the collisions can be understood and modeled properly. However, photons from jet plasma interaction dominates at high $p_T$ region where the tranverse expansion is negligible. Our primary concern here is to see the effect of pre-equilibrium momentum space anisotropy on the jet-photon production.

Photon (dilepton) production from relativistic heavy-ion collisions has been extensively studied in Ref. [12, 13, 14, 15]. All these works are based on the assumption of rapid thermalization of plasma with $\tau_{\text{therm}} = \tau_{\text{i}}$, where $\tau_{\text{i}}$ is the time scale of plasma formation. However, due to rapid longitudinal expansion of the plasma at early time this assumption seems to be very drastic because it ignores the momentum space anisotropy developed along the beam axis.

The phenomenological consequences of early stage pre-equilibrium momentum space anisotropy of QGP have been studied in Ref. [16, 17] in the context of dileptons and in Ref. [18, 19] in the context of photons. In Ref. [18] the effects of time-dependent momentum-space anisotropy of QGP on the medium photon production are discussed. It is shown that the introduction of early time momentum-space anisotropy can modify the photon production yield significantly. Also the present authors calculate transverse momentum distribution of direct photons from various sources by taking into account the initial state momentum anisotropy of QGP and the late stage tranverse flow effects [19]. The total photon yield is compared with the recent measurement of photon transverse momentum distribution by the PHENIX Collaboration [20] to extract the isotropization time [19]. It is found that the data can be reproduced with $\tau_{\text{iso}}$ in the range $0.5 - 1.5$ fm/c. All these works show that the introduction of pre-equilibrium momentum space anisotropy has significant effect on the medium dilepton as well as photon productions. In the present work, we will be investigating the $p_T$ distribution of the jet conversion photons in the presence of pre-equilibrium momentum space anisotropy.

Photons from jet conversion mechanism are produced when a high energy jet interacts with the medium constituents via annihilation and Compton processes [21]. It might be noted that this phenomenon (for Compton process) has been illustrated quite some time ago [22] in the context of estimating photons from equilibrating plasma, where, it is assumed that because of the larger cross-section, gluons equilibrate faster providing a heat bath to the incoming quark-jet. It is to be noted that, while evaluating jet conversion photons the assumption made in Ref. [21] that the largest contribution to photons corresponds to $p_\gamma \sim p_q(p_{\bar{q}})$ which implies that the annihilating quark (anti-quark) directly converts into a photon.

It is argued in Ref. [21] that measurement of photons from such a novel process can provide direct information about the quark momentum. This is because of the assumption made in Ref. [21] that photons are predominantly emitted at $p_\gamma \sim p_q$. This implies that the thermal distribution of the participating parton is evaluated at the photon momentum. In this work we relax this assumption, as it is shown that for LHC energies this is not a good approximation (exact calculation leads to a rate which is greater by a factor of $2-3$ [23]). We consider photon production in the $p_T$ range $4 \leq p_T \leq 15$ GeV.

In absence of any precise knowledge about the dynamics at early time of the collision, one can introduce phenomenological models to describe the evolution of the pre-equilibrium phase. In this work, we will use one such model, proposed in Ref. [16], for the time dependence of the anisotropy parameter, $\xi(\tau)$, and hard momentum scale, $p_{\text{hard}}(\tau)$. This model introduces four parameters to parameterize the ignorance of pre-equilibrium dynamics: the parton formation time ($\tau_i$), the isotropization time ($\tau_{\text{iso}}$), which is the time when the system starts to undergo ideal hydrodynamical expansion and $\gamma$ sets the sharpness of the transition to hydrodynamical behavior. The fourth parameter $\delta$ is introduced to characterize the nature of pre-equilibrium anisotropy i.e whether the pre-equilibrium phase is non-interacting or collisionally broadened.

The organization of the paper is as follows. In the next section (section 2) we shall discuss the mechanisms of jet conversion photon production rate using an anisotropic phase space distribution along with the space-time evolution of the matter. Results are presented in section 3 and finally we conclude in section 4.

## 2 Formalism

### 2.1 Jet conversion photon rate : Anisotropic QGP

The lowest order mechanisms for photon emission from QGP are the Compton ($q(\bar{q})\,g \rightarrow q(\bar{q})\,\gamma$) and the annihilation ($q\,\bar{q} \rightarrow g\,\gamma$) processes. The rate of thermal photon production from anisotropic plasma due to Compton and annihilation processes has been calculated in Ref. [24]. The soft contribution is calculated by evaluating the photon polarization tensor for an oblate momentum-space anisotropy of the system where the cut-off scale is fixed at $k_c \sim \sqrt{g} p_{\text{hard}}$. Here $p_{\text{hard}}$ is a hard-momentum scale that appears in the distribution functions. Apart from the thermal interactions of the plasma partons, interaction of a leading jet parton with the plasma was found to be a very important source of photons.

Now the differential cross-sections for Compton and annihilation processes are given by [25],

$$\frac{d\sigma(qg \rightarrow q\gamma)}{d\hat{t}} = \frac{1}{6}(\frac{e_q}{e})^2 \frac{8\pi\alpha_s\alpha_e}{(\hat{s}-m^2)^2}[(\frac{m^2}{\hat{s}-m^2} + \frac{m^2}{\hat{u}-m^2})^2$$
$$+ \ (\frac{m^2}{\hat{s}-m^2} + \frac{m^2}{\hat{u}-m^2}) - \frac{1}{4}(\frac{\hat{s}-m^2}{\hat{u}-m^2} + \frac{\hat{u}-m^2}{\hat{s}-m^2})] \quad (1)$$

and

$$\frac{d\sigma(q\bar{q} \rightarrow g\gamma)}{d\hat{t}} = -\frac{4}{9}\frac{8\pi\alpha_s\alpha_e}{\hat{s}(\hat{s}-4m^2)}[(\frac{m^2}{\hat{t}-m^2} + \frac{m^2}{\hat{u}-m^2})^2$$

$$+ \left(\frac{m^2}{\hat{t}-m^2}+\frac{m^2}{\hat{u}-m^2}\right) - \frac{1}{4}\left(\frac{\hat{t}-m^2}{\hat{u}-m^2}+\frac{\hat{u}-m^2}{\hat{t}-m^2}\right)] \tag{2}$$

where $m$ is the in-medium thermal quark mass. $m^2 = 2m_{th}{}^2 = 4\pi\alpha_s T^2/3$, $\alpha_e$ and $\alpha_s$ are the electromagnetic fine-structure constant and the strong coupling constant, respectively.

The differential photon production rate for these processes is given by:

$$\begin{aligned}E\frac{dR}{d^3p} &= \frac{\mathcal{N}}{2(2\pi)^3}\int \frac{d^3p_1}{2E_1(2\pi)^3}\frac{d^3p_2}{2E_2(2\pi)^3}\frac{d^3p_3}{2E_3(2\pi)^3} f_{jet}(\mathbf{p_1}) f_2(\mathbf{p_2},p_{\text{hard}}(\tau),\xi) \\ &\times (2\pi)^4\delta(p_1+p_2-p_3-p)|\mathcal{M}|^2[1\pm f_3(\mathbf{p_3},p_{\text{hard}}(\tau),\xi)]\end{aligned} \tag{3}$$

where, $|\mathcal{M}|^2$ represents amplitude for Compton or annihilation process and be obtained from Eqs. (1) and (2). $\mathcal{N}$ is the degeneracy factor of the corresponding process. $\xi$ is a parameter controlling the strength of the anisotropy with $\xi > -1$. $f_{\text{jet}}$, $f_2$ and $f_3$ are the distribution functions of the initial state and final state partons. Here it is assumed that the infrared singularities can be shielded by the thermal masses for the participating partons. This is a good approximation at times short compared to the time scale when plasma instabilities start to play an important role. The rate given in Eq. (3) is the static rate which has to be convoluted with the space-time history of the plasma (see section 2.2) to obtain phenomenologically predictable quantities. We have assumed that the plasma is formed at time (proper) $\tau_i$ with a temperature $T_i$ and it undergoes phase transition (at a critical temperature $T_c$) which begins at the time $\tau = \tau_f$ and ends at $\tau_H = r_d\tau_f$, where $r_d = g_Q/g_H$ is the ratio of the degrees of freedom in the two (QGP phase and hadronic phase) phases. With these assumptions, one obtains the space time integrated photon yield as,

$$\frac{dN}{dyd^2p_T} = \pi R'^2 \left[\int_{\tau_i}^{\tau_f}\tau d\tau \int d\eta \frac{dR}{dyd^2p_T} + \int_{\tau_f}^{\tau_H} f_{QGP}(\tau)\tau d\tau \int d\eta \frac{dR}{dyd^2p_T}\right], \tag{4}$$

where, $f_{\text{QGP}}(\tau) = (r_d-1)^{-1}(r_d\tau_f\tau^{-1}-1)$, is the fraction of the QGP phase in the mixed phase [26], $R' = 1.2[\langle N_{\text{part}}\rangle/2]^{1/3}$ fm is the dimension of the system, $\langle N_{\text{part}}\rangle$ is the average number of participants for a given centrality class. The energy of the photon in the fluid rest frame is given by $E_\gamma = p_T\cosh(y-\eta)$ where $\eta$ and $y$ are the space-time and photon rapidities respectively.

In the case of anisotropic QGP, the anisotropic distribution function can be obtained [27] by squeezing or stretching an arbitrary isotropic distribution function along the preferred direction in the momentum space,

$$f_i(\mathbf{p},\xi,p_{\text{hard}}) = f_i^{\text{iso}}(\sqrt{\mathbf{p}^2+\xi(\mathbf{p}.\mathbf{n})^2},p_{\text{hard}}(\tau)) \tag{5}$$

where $\mathbf{n}$ is the direction of anisotropy. It is important to notice that $\xi > 0$ corresponds to a contraction of the distribution function in the direction of anisotropy and $-1 < \xi < 0$ corresponds to a stretching in the direction of anisotropy. In the context of relativistic heavy ion collisions, one can identify the direction of anisotropy with the beam axis along which the system expands initially. The hard momentum scale $p_{\text{hard}}$ is directly related to the average momentum of the partons. In the case of an isotropic QGP, $p_{\text{hard}}$ can be identified with the plasma temperature ($T_i$).

Jet conversion photons arises from the electromagnetic interaction of jet-quarks, originated from the hard-scattering of colliding nuclei, with the plasma partons. The phase space distribution for the jet quarks is given by [21],

$$f_{jet}(\mathbf{p_1}) = \frac{1}{g_q} \frac{(2\pi)^3}{\pi R'^2 \tau p_1} \frac{dN_{jet}}{d^2 p_{1T} dy} R(r)$$
$$\times \ \delta(\eta - y) \ \Theta(\tau_f - \tau_i) \ \Theta(R' - r) \quad (6)$$

where, $g_q = 2 \times 3$ is the spin and colour degeneracy factor and $R(r)$ is the tranverse profile function [21]. The $p_T$ distribution of the jet-(anti)quarks and gluons in the central rapidity region ($y = 0$) was computed in Ref. [21] and parameterized as:

$$\frac{dN_{jet}}{d^2 p_{1T} dy}|_{y=0} = T_{AA} \frac{d\sigma_{jet}}{d^2 p_T dy}|_{y=0} = K \frac{a}{(1 + p_1/b)^c} \quad (7)$$

This is a tree-level result. To account the higher order effects, a $K$ factor has been introduced. The numerical values of the parameters $a$, $b$ and $c$ are taken from Ref. [21].

Using the Fermi-Dirac distribution function for the initial plasma parton and Bose Einstein distribution function for the final state gluon (for Compton scattering) or Fermi-Dirac distribution for the final state quark (for annihilation process), the jet conversion photon yield was computed previously in Ref. [21]. In this article, we will concentrate on the jet-photon production from an anisotropic QGP medium. Pre-equilibrium momentum space anisotropy could not affect the jet-quarks distribution function. However, the plasma parton distribution functions significantly depend on the amount of plasma anisotropy. Thus we could expect significant modification of jet-photon yield as a consequence of plasma anisotropy. In order to integrate Eq. 4 for an anisotropic medium, one needs to know the time dependence of $p_{\text{hard}}$ and $\xi$ which will be discussed in the following.

## 2.2 Space time evolution

The exact dynamics at the early-stage of the heavy ion collision is almost unknown. Thus, a precise theoretical picture of the evolutions of $p_{\text{hard}}$ and $\xi$ is not possible. However, we can always introduce phenomenological models to parameterize the ignorance. In this work, we shall closely follow the works of Refs. [16, 17, 18, 19] to evaluate the $p_T$ distribution of jet conversion photons from the first few Fermi of the plasma evolution. Three scenarios of the space-time evolution (as described in Ref. [17]) are the following: (i) $\tau_{\text{iso}} = \tau_i$, the system evolves hydrodynamically so that $\xi = 0$ and $p_{\text{hard}}$ can be identified with the temperature ($T$) of the system (till date all the calculations have been performed in this scenario), (ii) $\tau_{\text{iso}} \to \infty$, the system never comes to equilibrium, (iii) $\tau_{\text{iso}} > \tau_i$ and $\tau_{\text{iso}}$ is finite, one should devise a time evolution model for $\xi$ and $p_{\text{hard}}$ which smoothly interpolates between pre-equilibrium anisotropy and hydrodynamics. We shall follow scenario (iii) (see Ref. [17] for details) in which case the time dependence of the anisotropy parameter $\xi$ is given by

$$\xi(\tau, \delta) = \left(\frac{\tau}{\tau_i}\right)^\delta - 1 \quad (8)$$

where the exponent $\delta = 2$ (2/3) corresponds to *free-streaming (collisionally-broadened)* pre-equilibrium momentum space anisotropy and $\delta = 0$ corresponds to complete isotropization.

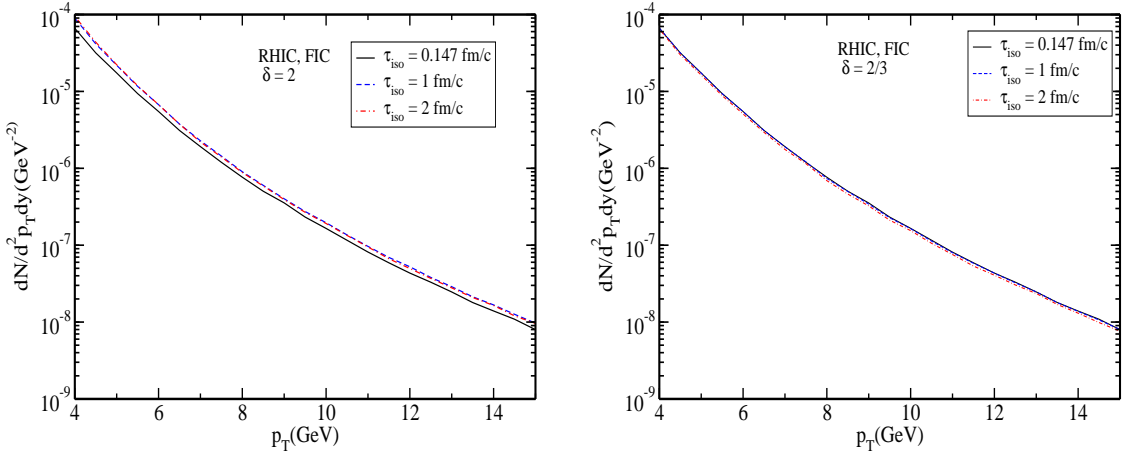

Figure 1: (Color online) $p_T$ distributions of jet conversion photons for FIC interpolating model at (a) $\delta = 2$ and (b) $\delta = 2/3$ at RHIC energy. Here $T_i = 0.446$ GeV and $\tau_{\text{iso}} = 0.147$ fm/c.

As in Ref. [17], a transition width $\gamma^{-1}$ is introduced to take into account the smooth transition from non-zero value of $\delta$ to $\delta = 0$ at $\tau = \tau_{\text{iso}}$. The time dependence of various quantities are, therefore, obtained in terms of a smeared step function [16]:

$$\lambda(\tau) = \frac{1}{2}(\tanh[\gamma(\tau - \tau_{\text{iso}})/\tau_i] + 1). \tag{9}$$

For $\tau \ll \tau_{\text{iso}} (\gg \tau_{\text{iso}})$ we have $\lambda = 0(1)$ which corresponds to *free streaming* (hydrodynamics). With this, the time dependence of relevant quantities are as follows [17, 28]:

$$\begin{aligned}
\xi(\tau, \delta) &= \left(\frac{\tau}{\tau_i}\right)^{\delta(1-\lambda(\tau))} - 1, \\
p_{\text{hard}}(\tau) &= T_i\, \bar{\mathcal{U}}^{1/3}(\tau),
\end{aligned} \tag{10}$$

where,

$$\begin{aligned}
\mathcal{U}(\tau) &\equiv \left[\mathcal{R}\left((\frac{\tau_{\text{iso}}}{\tau})^\delta - 1\right)\right]^{3\lambda(\tau)/4} \left(\frac{\tau_{\text{iso}}}{\tau}\right)^{1-\delta(1-\lambda(\tau))/2}, \\
\bar{\mathcal{U}} &\equiv \frac{\mathcal{U}(\tau)}{\bar{\mathcal{U}}(\tau_i)}, \\
\mathcal{R}(x) &= \frac{1}{2}[1/(x+1) + \tan^{-1}\sqrt{x}/\sqrt{x}]
\end{aligned} \tag{11}$$

and $T_i$ is the initial temperature of the plasma. After introducing the space-time evolution of different relevant quantities, we are now completely equipped for the integration of Eq. (4). However, before going into the details of $p_T$ distributions of jet conversion photons at the heavy ion collider experiments like RHIC and LHC, let us concentrate on few interesting features of the phenomenological model introduced for the evolution of $p_{\text{hard}}(\tau)$ and $\xi(\tau)$.

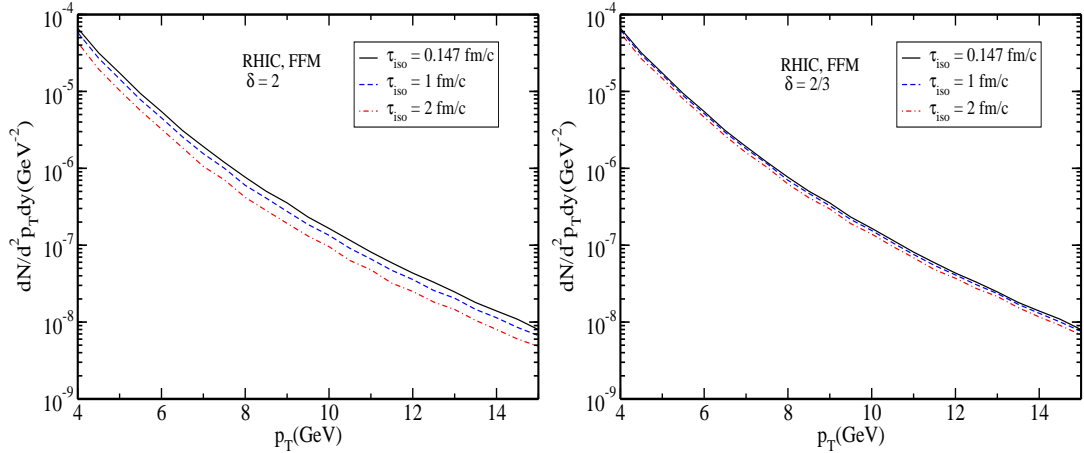

Figure 2: (Color online) $p_T$ distributions of jet conversion photons for FFM interpolating model at (a) $\delta = 2$ and (b) $\delta = 2/3$ at RHIC energy. Here $T_i = 0.446$ GeV and $\tau_{\text{iso}} = 0.147$ fm/c.

The simple model, introduced above, which smoothly interpolates between an initially non-equilibrium plasma to an isotropic plasma, is based on the assumption that the initial conditions are held fixed. The smooth interpolation (keeping the initial condition fixed) between anisotropic and isotropic phases (described in this section) results into a hard momentum scale (for $\tau \gg \tau_{\text{iso}}$) which is, by a factor $[\mathcal{R}((\tau_{\text{iso}}/\tau_i)^\delta - 1)]^{0.25}$, larger compared to the momentum scale results from the hydrodynamic expansion of a system (with the same initial condition) from the beginning (see Eqs. (10) and (11)). As a consequence of this enhancement of $p_{\text{hard}}(\tau)$, the *fixed initial condition* (FIC) interpolating models (both *free-streaming* and *collisionally-broadened*) will result in generation of particle number during the transition from non-equilibrium to equilibrium phase. Moreover, the entropy generation increases with the increasing value of $\tau_{\text{iso}}$. Thus, the requirement of bounded entropy generation can be used to put some upper bound on the value of $\tau_{\text{iso}}$ for *fixed initial condition* interpolating models [17].

Due to the phenomenological constraints on the entropy generation, one might not allow any entropy generation at all. In that case, one can redefine $\bar{\mathcal{U}}(\tau)$ in Eq. (10) to ensure *fixed final multiplicity* (FFM) in this model. Since we know the amount of enhancement (which is respondable for this entropy generation) of $p_{\text{hard}}$, the redefinition of $\bar{\mathcal{U}}(\tau)$ will be straight forward [17]:

$$\bar{\mathcal{U}}(\tau) = \mathcal{U}(\tau) \left[\mathcal{R}((\tau_{\text{iso}}/\tau_i)^\delta - 1)\right]^{-3/4} (\tau_i/\tau_{\text{iso}}) \tag{12}$$

It is important to mention that this redefinition corresponds to a lower initial temperature ($p_{\text{hard}}(\tau_i) < T_i$) for $\tau_{\text{iso}} > \tau_i$. Larger value of isotropization time corresponds to lower initial temperature.

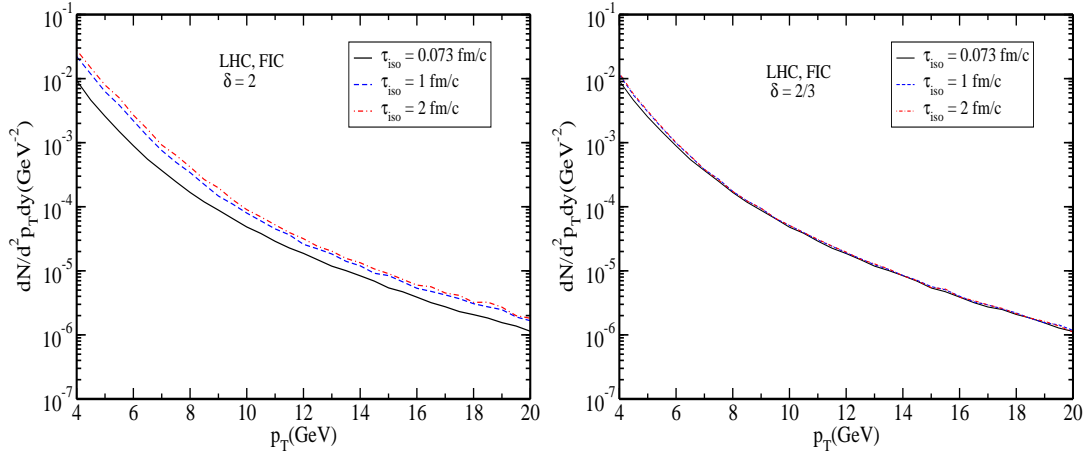

Figure 3: (Color online) $p_T$ distributions of jet conversion photons for FIC interpolating model at (a) $\delta = 2$ and (b) $\delta = 2/3$ at LHC energy. Here $T_i = 0.845$ GeV and $\tau_{\text{iso}} = 0.088$ fm/c

# 3 Results

## 3.1 Fixed initial condition (FIC) interpolating model

Fixed initial condition interpolating models always result into an enhanced value of hard momentum scale as a consequence of pre-equilibrium anisotropy. This feature of this model has already been discussed briefly before (see Ref. [17] for details). As a consequence of this enhancement of $p_{\text{hard}}$, pre-equilibrium anisotropy increases the density of plasma partons. The anisotropic parton distribution function in Eq. (5) clearly suggests that for the positive values of $\xi$, parton density decreases if we decrease the angle between the momentum of partons and the direction of anisotropy. From the very beginning, we have assumed that pre-equilibrium momentum-space anisotropy (in the heavy ion collisions) results from the rapid longitudinal expansion of the system immediately after the collision. Thus, in the context of relativistic heavy ion collision, one can identify the direction of anisotropy as the beam axis. Moreover, due to the rapid longitudinal cooling (as a consequence of rapid longitudinal expansion), one always finds oblate anisotropic distributions i.e positive value of $\xi$. Therefore, introduction of pre-equilibrium anisotropy with fixed initial condition increases the density of plasma partons moving in the transverse direction [18] and at the same time decreases the density of plasma partons moving in the forward direction. This feature of fixed initial condition pre-equilibrium momentum-space anisotropy should be reflected in the jet conversion photon $p_T$ distribution which we will see in the following.

As a consequence of this enhancement of partons moving in the transverse direction and suppression of partons moving in forward direction, we expect an enhancement in the transverse direction of the yield $(y = 0)$[4]. The enhancement is not uniform. This feature is better understood by looking at Fig. 1. In Fig. 1, we have presented the $p_T$ distribution

---

[4]This feature was already established in our previous work [18].

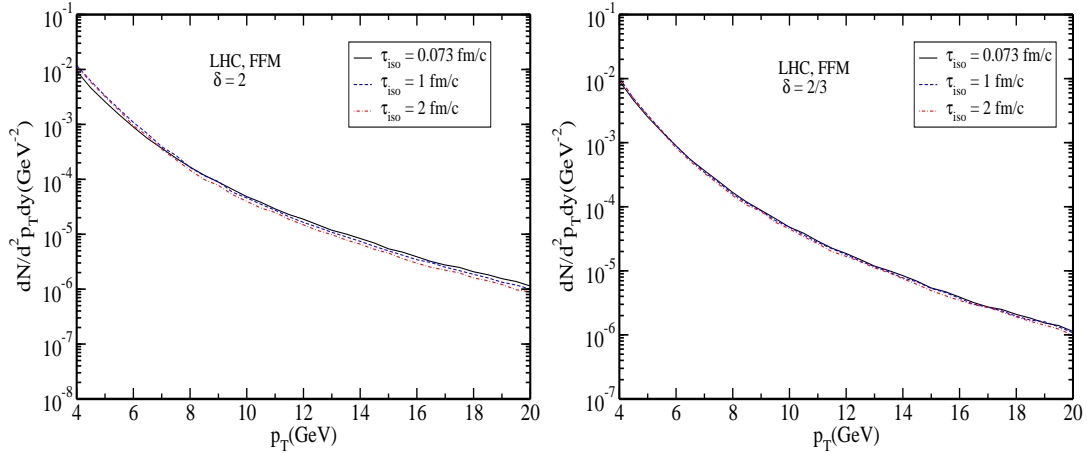

Figure 4: (Color online) $p_T$ distributions of jet conversion photons for FFM interpolating model at (a) $\delta = 2$ and (b) $\delta = 2/3$ at LHC energy. Here $T_i = 0.845$ GeV and $\tau_{\text{iso}} = 0.088$ fm/c

of jet conversion photons (for three different values of isotropization time, $\tau_{\text{iso}} = \tau_i$, 1 and 2 fm/c) in the framework of *fixed initial condition* for (a) *free-streaming* ($\delta = 2$) and (b) *collisionally-broadened* ($\delta = 2/3$) interpolating models.

Fig. 1b shows that for *fixed initial condition collisionally-broadened* interpolating model, the enhancement in the transverse direction is small compared to the *free-streaming* model (see Fig. 1a). This can be attributed to the fact that in the case of *collisionally broadened interpolating* model we have included the possibility of momentum space broadening of the plasma partons due to interactions. As a consequence the hard momentum scale $p_{\text{hard}}$ (which is related to the average momentum in the partonic distribution functions) decreases with time, whereas for *free streaming* model, the hard momentum scale remains unchanged ($p_{\text{hard}}(\tau) = p_{\text{hard}}(\tau_i) = T_i$, for $\tau < \tau_{\text{iso}}$) upto $\tau = \tau_{\text{iso}}$.

Fig. 1 also shows the enhancement of the jet conversion photon yields in the whole $p_T$ range ($4 \le p_T \le 15$). More enhancement is observed for $\delta = 2$ compared to $\delta = 2/3$. These findings are similar to that in Ref. [18] where for $\delta = 2$ with FIC considerable enhancement is observed at $y = 0$. But for $\delta = 2/3$, it is found that the enhancement is small.

## 3.2 Fixed final multiplicity (FFM) interpolating model

The problems, regarding the entropy generation (appears in *fixed initial condition* model), could be eliminated by enforcing *fixed final multiplicity*. However, enforcing *fixed final multiplicity* corresponds to a lower value of initial hard momentum scale ($p_{\text{hard}} < T_i$). Moreover, in this case, the initial hard momentum scale will be isotropization time dependent i.e larger the value of $\tau_{\text{iso}}$ lower will be the initial hard momentum scale. The suppression of hard momentum scale corresponds to a suppression in the plasma parton density compared to the *fixed initial condition* interpolating model. Therefore, for *fixed final multiplicity* interpolating model, we find suppression in the jet conversion photon yield. As a consequence of non-zero positive anisotropy parameter $\xi$, the yields of jet conversion photons will be more suppressed.

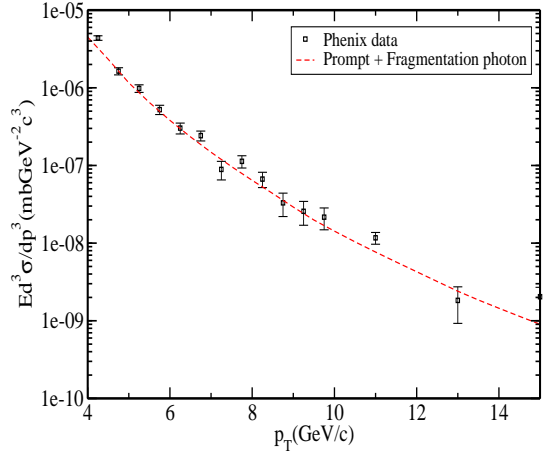

Figure 5: (Color online) Prompt and fragmentation Photons compared with PHENIX $p-p$ data [31].

In Fig. 2 we have plotted the $p_T$ distribution of jet conversion photons in the framework of *fixed final multiplicity* for (a) *free streaming* ($\delta = 2$) and (b) *collisionally-broadened* ($\delta = 2/3$) interpolating models. Both Figs. 2a and 2b show suppression of the jet conversion photon yield with increasing $\tau_{\text{iso}}$.

For $\delta = 2/3$ the rate is marginally suppressed. However, for $\delta = 2$ substantial suppression is observed for the entire $p_T$ range. This is due to the fact that *collisionally-broadened* interpolating model is always closer to the local-isotropic expansion.

In Fig. 3 we have plotted the yields of jet conversion photons as a function of transverse momentum ($p_T$) in the framework of *fixed final multiplicity* for (a) *free streaming* ($\delta = 2$) and (b) *collisionally-broadened* ($\delta = 2/3$) interpolating models at LHC energy. Both Figs. 3a and 3 b show enhancement of the jet conversion photon yield with increasing $\tau_{\text{iso}}$ as in the case of the RHIC.

Similarly, the $p_T$ distribution of jet conversion photons for FFM interpolating model is shown in Fig. 4 for $\sqrt{S_{NN}} = 5.5$ TeV. We find suppression of the jet conversion photon yield for both $\delta = 2$ (Fig. 4a) and $\delta = 2/3$ (Fig. 4b) for the reason mentioned above.

Let us now mention the main features of our observations. With FIC for both values of $\delta$ we see moderate enhancement. Similar feature has been noted earlier [18]. This can be attributed to the fact that momentum anisotropy enhances the density of plasma partons in the tranverse direction. This is similar to the findings in Ref. [28] for the case of dilepton. It is worthwhile to mention that the effects of pre-equilibrium momentum space anisotropy for $\delta = 2$ is more compared to that for $\delta = 2/3$. This is because $\delta = 2/3$ corresponds to close to isotropization. With FFM we observe considerable suppression for both free-streaming and collisionally broadened interpolating models, similar to the observation in Ref. [18, 19, 28]. This suppression can be described in two ways. Due to rapid longitudinal expansion the distribution function becomes anisotropic. Jet conversion photons with the larger values of longitudinal momentum are reduced compared with the jet conversion photons with isotropic distribution function. Maximum amount of momentum-space anisotropy achieved in the early times will be the important cause of the suppression. The suppression will also depend on the time dependence of the anisotropy parameter $\xi$.

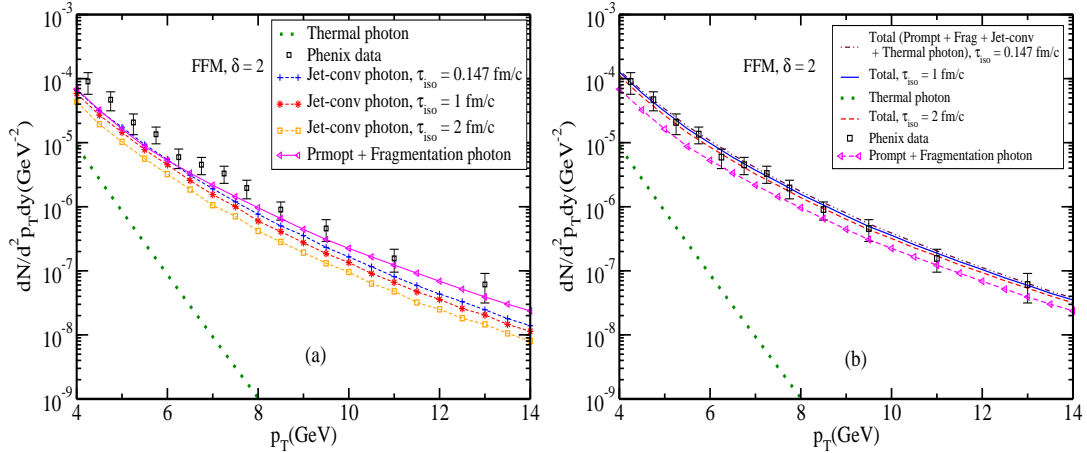

Figure 6: (Color online) Comparison with Phenix $A-A$ data. Here $T_i = 0.446$ GeV, $\tau_i = 0.147$ fm/c. (a) Individual contributions are shown for various $\tau_{iso}$, (b) Total contributions including prompt and fragmentation photons for various $\tau_{iso}$.

Apart from jet conversion photons, there are other non-thermal sources of photon like, prompt photons, fragmentation photons, decay photons etc. Prompt photons are produced in the initial hard scattering of colliding nuclei and are insensitive to the later stage evolution of the QGP. Therefore, they do not carry any information about the pre-equilibrium anisotropy. Similarly, decay photons are also not sensitive to the pre-equilibrium phase.

It is to be mentioned that apart from thermal photons from quark matter thermal photons are also produced from hot hadronic matter where the late stage transverse expansion plays important role. However, these are also insensitive to the pre-equilibrium momentum space anisotropy and contribute at lower $p_T$. Hence we do not consider it here. The importance of jet conversion photon in the context of PHENIX [29] photon data was shown in Ref. [15]. For jet-plasma interaction, the jet-parton distribution functions do not depend on the pre-equilibrium phase. However, the plasma parton densities depend on the pre-equilibrium anisotropy. Therefore, the jet-plasma interaction is sensitive to the pre-equilibrium anisotropy. Moreover, the jet conversion photon contribution dominates over the thermal photon contribution only in the high $p_T$ region.

We calculate the prompt photons along with fragmentation photons using the method described in [30] in LO. The Phenix $p-p$ data at $\sqrt{s} = 200$ GeV [31] is compared in Fig. 5. To take into account higher order processes we use so called K factor. It is seen that the data is well reproduced when $K = 2.5$. Once the parameters are fixed by comparing with $p-p$ data [31] we can now proceed for $A-A$ collisions. The yield is given by

$$\frac{dN}{d^2p_T dy} = \langle T_{AB}(b) \rangle \frac{d\sigma}{d^2p_T dy} \qquad (13)$$

where $\langle T_{AB}(b) \rangle = 21.7 mb^{-1}$ for $0-10\%$ centrality.

The total photon yield calculated (at RHIC energy with $T_i = 0.446$ GeV, $\tau_i = 0.147$ fm/c, $\delta = 2$ in FFM interpolating model) for various sources are plotted along with the PHENIX photon data [29] in Fig. 6a. It is observed that the contribution of jet-conversion photons for isotropic QGP ($\tau_{\text{iso}} = 0.147 fm/c$) is of the same order as prompt plus fragmentation contribution upto $p_T \sim 7$ GeV. But for anisotropic QGP the former always remains below the prompt and fragmentation photons (see the graphs for $\tau_{\text{iso}} = 1$ and $2$ $fm/c$ in Fig.6a). We also observe here that the data cannot be explained by the prompt and fragmentation contributions alone specially in the $p_T$ range $4 \le p_T \le 6$ GeV (see Figs.6a and 6b). Thus it is essential to consider the contribution of jet conversion photons to explain the data as has already been argued in [15]. Fig. 6b shows the total yield for three values of $\tau_{\text{iso}}$. Here we demonstrate that the presence of such an anisotropy can describe the PHENIX photon data [29] quite well. It is found that the data is better described if $\tau_{\text{iso}} \le 1.5$ fm/c which is similar to the findings in Ref. [19]. It is important to note here that the fragmentation photons are sensitive to the pre-equilibrium momentum space anisotropy in the sense that the partons before fragmenting to photons lose energy in the medium due to collisional and radiative processes. The collisional energy loss in anisotropic QGP has been estimated in Ref. [32] while the radiative energy loss in anisotropic media is under investigation [33]. We do not include the effect of energy loss in the photon yield both for jet-conversion and fragmentation photons. Inclusion of this effect will reduce the yields in both the cases.

We note that there are other observables (non-photonic) where the present model can be applied. For example, the non-photonic single electron (from the semi-leptonic decay of heavy quarks) $p_T$ spectra (or nuclear modification factor, $R_{AA}$) can be one such probe. To estimate this one must calculate heavy quark energy loss due to collisional and radiative processes in anisotropic medium. Also the space-time evolution in the anisotropic medium will depend on the isotropization time, $\tau_{\text{iso}}$. One can also apply this model to investigate $R_{AA}$ of light hadrons to see the importance of pre-equilibrium momentum space anisotropy. Like the photonic observables it is also possible to extract $\tau_{\text{iso}}$ from these probes and one can check whether $\tau_{\text{iso}}$ extracted from photon signals matches with that obtained by comparing with non-photonic observables.

It is also important to recall that RHIC data in semi-central collisions exhibit a strong elliptic flow which indicates rapid thermalization of the matter produced in the collisions [34]. On the contrary, perturbative estimation suggests the slower thermalization of the QGP [35]. It is suggested that (momentum) anisotropy driven plasma instabilities may speed up the process of isotropization [36]. If we assume that the latter is true, it will also be interesting to calculate the elliptic flow ($v_2$) of heavy quark with momentum space anisotropy in semi-central collisions. To calculate $v_2$ in anisotropic medium one needs to calculate drag and diffusion coefficients that enter in the FP equation which again, will depend on the parameter $\tau_{\text{iso}}$ through space-time dynamics. Thus, in this case, it is also possible to extract this parameter and see whether it matches with that obtained from other observables. All these probes for anisotropic QGP are left for future investigations.

# 4   Conclusion

To summarize, we have investigated the effects of the pre-equilibrium momentum space anisotropy of the QGP on the $p_T$ distribution of the jet conversion photons. To describe space-time evolution of hard momentum scale, $p_{\text{hard}}(\tau)$ and anisotropy parameter, $\xi(\tau)$, two phenomenological models have been used [17]. These phenomenological models assume the existence of an intermediate time scale called the isotropization time ($\tau_{\text{iso}}$). The first model is based on the assumption of fixed initial condition. However, enforcing fixed initial condition causes entropy generation. Therefore, we have also considered another model, which assumes the fixed final multiplicity. The $p_T$ distribution of the jet conversion photons for different isotropization times in the frame work of these phenomenological models have been estimated. We observed that, for fixed initial condition, a *free streaming* interpolating model can enhance the jet conversion photon yield significantly. However, for *collisionally broadened* pre-equilibrium phase with fixed initial condition, the enhancement of photon yield is not that much. Since fixing the final multiplicity reduces the initial hard momentum scale or equivalently the initial energy density, the jet conversion photon yield is significantly suppressed (both for the *free streaming* and *collisionally-broadened* interpolating models). These findings with FFM and $\delta = 2/3$ are similar to that obtained for dilepton [18, 19, 28]. This suppression can be explained as a consequence of the combined effect of the anisotropy in momentum-space achieved at early times due to expansion. We note that although, the contribution of jet conversion photons always remains below that from prompt plus fragmentation photons in anisotropic plasma, the latter cannot explain the PHENIX photon data alone. The inclusion of jet-conversion photons from anisotropic QGP improves the agreement with the experimental data and thus constrains the isotropization time $\tau_{\text{iso}}$.

The other observables like heavy-quark transport [37] and gluonic radiation could be phenomenologically very useful in order to detect the consequences of pre-equilibrium momentum-space anisotropy. It is to be noted that the jet parton loses energy before producing photons. As a result the jet conversion photon yield will be lower and this will further constrain $\tau_{\text{iso}}$. However, we wish to include this effect in a separate work [38].